# Topological Valley-Reshaped Device: Bifunctional Waveguiding and Single-Beam Leaky-Wave Radiation for Terahertz Communication

Yulun Wu[1,2,3] | Ziwei Wang[4] | Faqian Chong[1,2,3] | Hua Shao[1,2,3] | Bingtao Gao[1,2,3] | Shilong Li[1,2,3] | Jin Tao[4,5] | Hongsheng Chen[1,2,3,6] | Song Han[1,2,3,7]

[1]Innovative Institute of Electromagnetic Information and Electronic Integration, College of Information Science & Electronic Engineering, Zhejiang University, Hangzhou, China | [2]State Key Laboratory of Extreme Photonics and Instrumentation, ZJU-Hangzhou Global Scientific and Technological Innovation Center, Zhejiang University, Hangzhou, China | [3]International Joint Innovation Center, The Electromagnetics Academy at Zhejiang University, Zhejiang University, Haining, China | [4]State Key Laboratory of Optical Communication Technologies and Networks (OCTN), China Information Communication Technologies Group Corporation (CICT), Wuhan, China | [5]Peng Cheng Laboratory, Shenzhen, China | [6]Key Lab. of Advanced Micro/Nano Electronic Devices & Smart Systems of Zhejiang, Jinhua Institute of Zhejiang University, Zhejiang University, Jinhua, China | [7]Zhejiang Key Laboratory of Advanced Micro-nano Transducers Technology, College of Information Science and Electronic Engineering, Zhejiang University, Hangzhou, China

**Correspondence:** Hongsheng Chen (hansomchen@zju.edu.cn) | Song Han (song.han@zju.edu.cn)

Yulun Wu and Ziwei Wang contributed equally to this work.

**ABSTRACT**

Topological photonics has emerged as a powerful platform for terahertz on-chip systems due to its robust waveguiding capabilities. However, directly extracting topological valley-locked edge states into directional free-space radiation—without auxiliary couplers while preserving guided-wave functionality—remains a fundamental challenge. In this work, we propose and experimentally demonstrate a bifunctional topological valley-reshaped device. By introducing an angular truncation and a spatial displacement to a complete topological waveguide (TW), the resulting structure inherently retains its waveguiding capabilities. Furthermore, when operated as an isolated section, it functions as a topological leaky-wave antenna (TLWA) that exhibits directional single-lobe radiation. The TW shows low-loss guided-wave performance with an 18 GHz operating bandwidth, supporting error-free transmission up to 60 Gbps. For the TLWA, by gradually reducing the number of protective lattices that are orthogonal to the propagation direction, the valley-locked edge state becomes momentum-matched to the free-space light line, generating leaky-wave radiation. Simultaneously, reshaping of the opposite valley-locked edge state suppresses far-field side lobes and reduces reflection, yielding a clean single-beam radiation pattern with side-lobe suppression ratio (SLSR) exceeding 15 dB. The TLWA realizes a measured peak gain of 12.5 dBi and a 19 GHz operating bandwidth. Notably, the low-dispersion property of the K-valley radiation allows the main-lobe direction varies by only 2° across the entire operating band, enabling error-free free-space reception at 24 Gbps. This bifunctional design represents a key step toward highly integrated and modular terahertz on-chip systems.

# 1 | Introduction

Terahertz wave offers broad spectral resources, high data rates, and low latency, making it a promising enabler for future wireless communication systems [1-4]. However, conventional metal-based waveguides and microstrip circuits [5-7] suffer from severe ohmic losses and fabrication challenges at terahertz frequencies, posing fundamental bottlenecks for on-chip integration. The emergence of topological valley photonics has largely mitigated these limitations in signal routing: valley-locked edge states exhibit strong robustness against disturbance and negligible bending loss even on the metal-based platform [8-10], and have been successfully implemented in various silicon-based terahertz devices, including waveguide [11-13], coupler [14, 15], switch [16,17], (de)multiplexer [18-20], beamformer [21], and resonant cavity [22-24].

Despite these advances in guided-wave transmission, extracting valley-locked edge states into directional free-space radiation in the terahertz band remains an open challenge—particularly without auxiliary couplers. Previous work has demonstrated coupling from valley-locked edge states to free space, but the resulting radiation either lacks directionality [25, 26] or requires additional structures such as taper [21] or graded refractive index buffer (GRIB) [27-29] to form a directional single-beam. Existing add-ons, though functional, introduce extra coupling loss and increase fabrication complexity and compromise mechanical robustness—posing critical barriers to large-scale integration and high-yield fabrication. Nevertheless, while certain metallic topological platforms have achieved directional radiation without auxiliary structures [30, 31], their operation remains confined to the millimeter-wave or even microwave regime and has yet to be extended to terahertz frequencies due to prohibitive ohmic losses and fabrication challenges [32, 33]. This highlights the need for an all-silicon, coupler-free solution operating in the terahertz band.

More importantly, whether relying on auxiliary structures or not, none of these antenna configurations can be seamlessly reintegrated into a waveguide to resume low-loss on-chip transmission. For instance, the leaky edge state can be directly coupled to free space [25]. Such an edge state resides above the light line and is inherently radiative, rendering it incapable of functioning as a guided-wave component. Notably, the chiral edge state has addressed this issue by enabling both evanescent-wave coupling to free space and on-chip waveguiding within the same topological device [34]. However, as experimentally verified, its far-field radiation capability remains extremely weak, failing to produce a well-defined beam with appreciable realized gain [29]. This fundamental limitation precludes the use of chiral edge state as standalone antennas.

Achieving coupler-free, highly-directional single-beam radiation directly from the valley-locked edge state while preserving the ability to revert to guided-wave operation is therefore highly desirable yet challenging. Such a capability would enable truly low-loss, highly integrated, and modular terahertz on-chip interconnected systems and significantly improve the fabrication yield of silicon-based photonic devices [35-36].

In this paper, we propose a topological valley-reshaped device that inherently possesses bifunctionality: derived from a complete topological waveguide (TW) via truncation at a specific angle, the device inherently features guided-wave function serving for the TW while independently functioning as a topological leaky-wave antenna (TLWA) without any additional structures. Experimental results show that the TW exhibits low-loss guided-wave performance with an 18 GHz operating bandwidth and supports error-free transmission up to 60 Gbps. When the TLWA is separated from the TW by designing special truncation angle, its original valley-locked edge state naturally transitions into a leaky mode, giving rise to directional radiation. This radiation is enabled by gradually reducing the number of protective lattices orthogonal to the propagation direction, achieving momentum-matching between the valley-locked edge state and the free-space light line. Meanwhile, reshaping of the opposite valley-locked edge state simultaneously suppresses far-field side lobes and reduces reflection, yielding a clean single-beam radiation pattern with side-lobe suppression ratio (SLSR) exceeding 15 dB. Owing to its unique K-valley leakage direction, the TLWA exhibits low dispersive behavior. Therefore, the main-lobe direction varies by only 2° across the entire operating band, enabling a 19 GHz operating bandwidth and a measured peak realized gain of 12.5 dBi. In the communication demonstration, the TLWA serves as the receiver and achieves error-free transmission at 24 Gbps. This bifunctional device, capable of both high-speed on-chip waveguiding and efficient free-space radiation, establishes a new strategy for highly integrated and modular terahertz transceivers.

## 2 | Results

### 2.1 | Design Principle

To enable effective coupling with WR-2.8 waveguide in subsequent simulations and experiments for terahertz operation, the lattice constant is chosen as $a$ = 242.5 μm, and the unit cell is patterned on lossless silicon wafer ($n$ = 3.45) with thickness ~200 μm, as shown by the dark blue rhombus dashed frame in Figure 1a. Based on this lattice, the silicon-based topological photonic waveguide is designed to form an AB-type interface, created by joining two mirror-symmetric photonic crystal domains at the sites of the larger equilateral triangular air holes with side length of $l_1$, as shown in Figure 1a. This configuration supports AB-type valley-locked topological edge state localized at the interface. The electromagnetic field is confined laterally along the domain wall, indicating robust interface-guided propagation, as illustrated in Figure 1c$_1$. Correspondingly, the transmission spectrum exhibits a highly efficient and flat-top passband within the topological bandgap, as represented by the yellow curve in Figure 1b.

To construct auxiliary-coupler-free radiation, the AB-type waveguide is divided into two separate sections by introducing different truncation angles, as shown in Figure 1c$_2$–c$_4$. Three representative truncation angles—0°, 30°, and 60°—are investigated. The two resulting sections are then horizontally displaced by a fixed relative distance for all cases in order to eliminate influence of free-space attenuation on the transmission comparison. Figure 1b compares the transmission spectra of the original AB-type waveguide with those of the split configurations for the three truncation angles. It can be observed that only the 60° truncation yields a transmission spectrum that closely reproduces that of the original AB-type waveguide. This result indicates that efficient leaky-wave coupling occurs under the 60° truncation condition, which enables the realization of highly directional and high-gain leaky-wave antennas without the need for auxiliary couplers.

The failure of the 0° and 30° truncation cases can be understood in light of previous studies [28, 29, 34]. At these truncation angles, the AB-type edge state cannot satisfy the momentum-matching condition with free space. Based on these observations, we focus our investigation on the 60° truncation configuration (Figure 1c4).

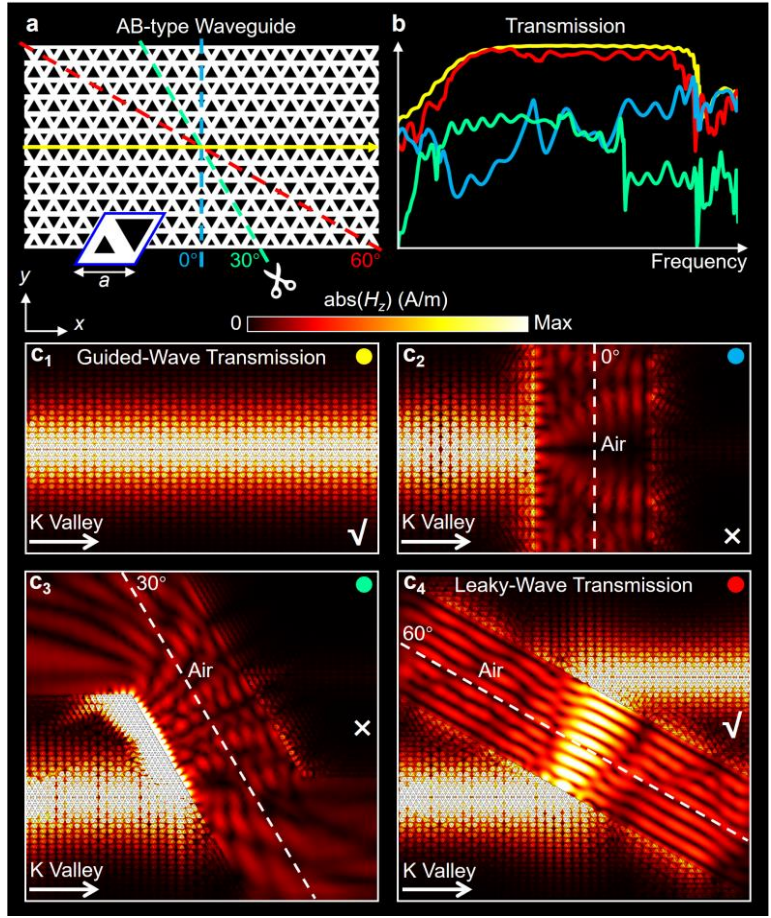

**FIGURE 1** | Simulated results of AB-type waveguide and its split configurations under the three truncation angles. (a) Top view of the AB-type waveguide and demonstration of different truncation angles. (b) Simulated transmission spectra of AB-type waveguide and its split configurations under different truncation angles. (c1-c4) Simulated magnetic field distributions of AB-type waveguide and its split configurations under different truncation angles.

As illustrated in Figure 2a, the wavevectors under the 60° truncation configuration can be explicitly depicted in real space. Here, $k_l$ denotes the projection of the K-valley momentum $k_{VPC}$ onto the guided-wave direction. When mapped into reciprocal space, all wavevector directions remain unchanged, as shown in Figure 2b. It can be observed that the radiation direction $k_{air}$ is consistent with the direction of $k_{VPC}$ of the AB-type edge state, and both are perpendicular to the truncation direction $e_t$. This provides a clear explanation from the perspective of momentum-matching condition for the realization of leaky-wave radiation (for a detailed derivation, please refer to Supporting Information S1).

To further elucidate the underlying mechanism, we simulate the band diagram of the AB-type edge states, as presented in Figure 2c. As the number of protective lattices $N$ decreases, the band of the AB-type edge state within the topological bandgap gradually approaches linear dispersion (dark red solid line). Consequently, the nearly linear band dispersion becomes phase-matched to the light line through evanescent-wave coupling. This further confirms the leaky-wave radiation condition from the perspective of band dispersion engineering.

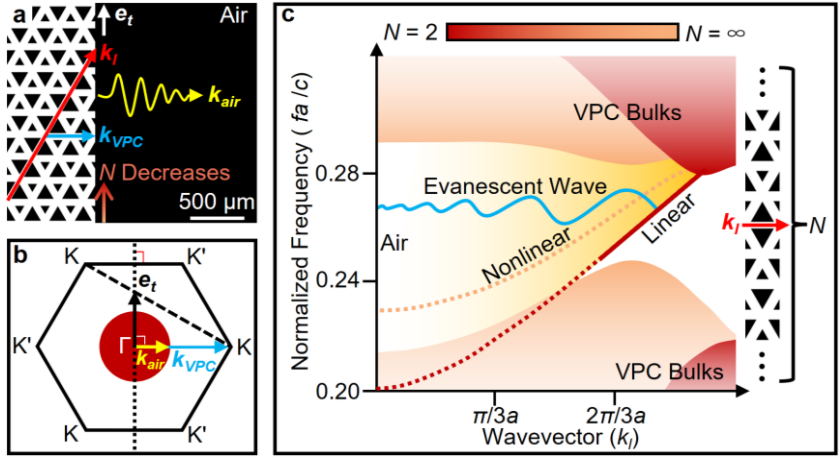

**FIGURE 2** | Momentum-matching condition and K-valley radiation (a) Top view of the leaky part of the 60° truncation configuration. $k_{VPC}$ is the K-valley momentum, $k_l$ is the direction of guided-wave, $e_t$ is the truncation direction, $k_{air}$ is the leaky-wave direction and $k_l$ is the projection of $k_{VPC}$ onto the guided-wave direction. (b) Demonstration of the wavevectors in the reciprocal space. (c) Band diagram of AB-type edge state under infinitely many protective lattices (light orange curve) and AB-type edge state under two protective lattices (dark red curve). The AB-type edge state under two protective lattices couples to air via evanescent wave (light blue curve).

## 2.2 | Global Device Design and Performance

Based on the leaky-wave radiation mechanism elucidated above, one might intuitively expect that a high-performance leaky-wave antenna could be realized simply by coupling a silicon-based taper coupler directly to an AB-type waveguide under the 60° truncation configuration. However, satisfying the momentum-matching condition between the K valley and free space alone is insufficient to address critical issues such as excessive reflection and pronounced side lobes. Consequently, while this configuration serves as the leaky-wave section, we introduce a BA-type waveguide as a preceding guided-wave section. The BA-type interface is created by joining two mirror-symmetric photonic crystal domains at the sites of the smaller equilateral triangular air holes with side length $l_2$. By directly coupling the taper coupler to this BA-type waveguide, the resulting phase-matching enables a reflection power below 1% and a side-lobe suppression level (SLSR) exceeding 15 dB, as shown in Figure 3a and Figure 3e$_1$, thereby achieving excellent performance in both guided-wave and radiation functionalities.

Notably, the frequency range covered by the BA-type edge state within the topological bandgap differs from that of the AB-type edge state. To address this, we assign distinct $\delta = l_1 - l_2$ values to the guided-wave section ($\delta = 0.25a$) and leaky-wave section ($\delta = 0.3a$), aiming to maximize the effective bandwidth—defined as the overlap between the impedance bandwidth and the transmission or radiation bandwidth. Detailed design considerations regarding $\delta$ are provided in Supporting Information S2. Ultimately, by integrating momentum-matching, phase-matching, and bandwidth-matching, the proposed device with low reflection and single-beam radiation characteristics within a large operating band is realized, as illustrated in Figure 3d$_1$.

For comparison, DEV. I employs the AB-type edge state for both guided-wave and leaky-wave sections, exhibiting a high reflection power of 30% and pronounced parasitic side lobes in the far field. DEV. II retains the AB-type edge state excited directly by the taper coupler but introduces a

BA-type edge state along the propagation path via two successive valley-locked edge-state conversions, its reflection power is reduced to 10%, yet parasitic side lobes persist. This contrast demonstrates that the reshaping capability inherent to valley-locked topological edge states is key to achieving simultaneous suppression of reflection and side lobes—a critical advantage for high-performance device applications.

The progressive reduction of the reflection coefficient can be explained from the perspective of dispersion engineering. As shown in the simulated results of Figure 3a, the proposed device maintains a reflection coefficient below −20 dB across its operating band—an achievement enabled by the distinct dispersion properties of the AB-type and BA-type edge states. Figure 3b reveals that the AB-type edge state possesses positive group velocity, whereas the BA-type edge state exhibits negative group velocity. Consequently, forward propagation enables efficient coupling from the AB-type to the BA-type edge state at the K valley. Conversely, backward propagation forbids coupling from the BA-type to the AB-type edge state at the K' valley. Exploiting this special coupling characteristic, cascading different valley-locked edge states effectively suppresses reflections caused by impedance mismatch at the silicon–air interface. Moreover, when the taper coupler is directly coupled to the BA-type edge state, the reflection coefficient is further reduced to a negligible level.

The single-beam radiation is enabled by the side-lobe suppression capability of BA–AB edge state coupling. To elucidate this mechanism, we simulate the phase distribution and its gradient $\beta_l = d(\text{phase})/dl$ along the lattice defect direction, the results are presented in Figure 3e$_1$–e$_3$. For DEV. I and DEV. II, $\beta_l$ exhibits continuous values within the 0.2 to 0.53 deg/μm range. In contrast, only the BA–AB valley-locked edge state coupled configuration (i.e., the proposed device) yields no solution for $\beta_l$ in this critical interval. The far-field normalized intensity derived in Supporting Information S3 is given below:

$$I(\theta,\varphi) = \left| \sum_{i=1}^{n} C_i \text{sinc}\left( \frac{\sqrt{3}D_1}{4} \beta_{l,i} - \frac{D_1}{2} k_0 \sin\theta \sin\varphi \right) \right|^2 \qquad (1)$$

In equation (1), $\theta$ and $\varphi$ are the polar angle and azimuth angle, respectively, $\beta_{l,i}$ refers to the $i$-th phase gradient, $D_1$ denotes the in-plane radiation aperture dimension of the proposed device (its specific dimension will be mentioned in the later section), $k_0$ is the momentum magnitude in free space, and $C_i$ refers to the weight of each far-field magnitude defined by each phase gradient. Substituting the obtained $\beta_l$ values into (1) rigorously demonstrates that parasitic side lobes are fully suppressed only when $\beta_l$ avoids the 0.2 to 0.53 deg/μm interval. Supporting Information S3 further provides detailed derivation of equation (1) and we can also draw an important conclusion from it: no matter how many times the valley-locked edge state switches during the guided-wave process, the final $\beta_l$ distribution depends only on the first valley-locked edge state. Therefore, the far-field radiation pattern can only be reshaped when the first valley-locked edge state is BA-type edge state.

Furthermore, by truncating the AB-type straight waveguide shown in Figure 1 along both the 60° and −60° directions and incorporating valley-reshaping mentioned above, dual-beam leaky-wave radiation with clean main lobes can also be achieved (for detailed simulated results of this modified TLWA, please refer to Supporting Information S4).

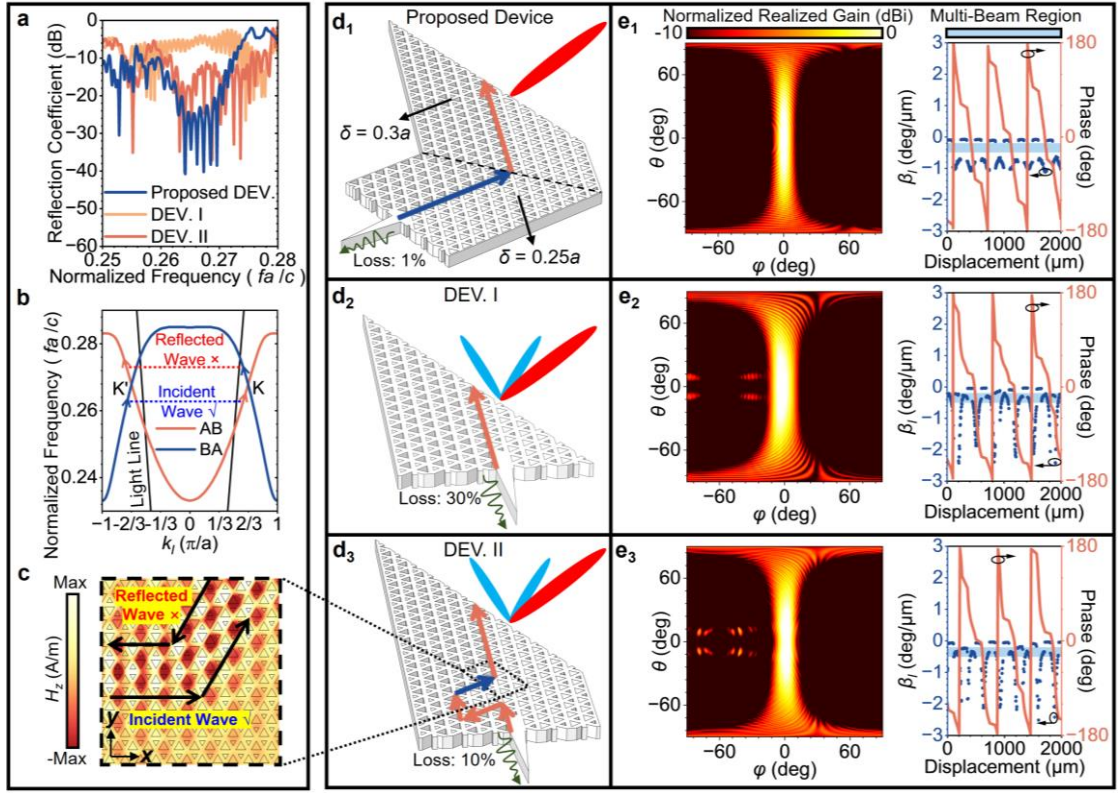

**FIGURE 3** | Presentation of the performance and the relevant details of the proposed device and other designs. (a) Simulated reflection coefficients of devices mentioned in the comparison. (b) Band diagram of the AB-type edge state and BA-type edge state. (c) Simulated magnetic field distribution in the coupling region between AB-type edge state and BA-type edge state. ($d_1$-$d_3$) The demonstrations of devices mentioned in the comparison. ($e_1$-$e_3$) Simulated far-field normalized radiation patterns and near-field phase and phase gradient distributions of devices mentioned in the comparison.

## 2.3 | Experimental Characterization

Based on the 60° truncation configuration derived from an AB-type waveguide, the proposed device can also be obtained from a complete topological waveguide (TW). As illustrated in Figure 4a, truncating the TW along the white dashed line yields two identical devices, each of which independently functions as a topological leaky-wave antenna (TLWA). In the following sections, we will experimentally characterize the guided-wave performance of the TW and the radiation performance of the TLWA.

The reflection coefficient and transmission spectrum of the TW are measured using a vector network analyzer (VNA), with the experimental setup illustrated in Figure 4a and the results shown in Figure 4b. The TW exhibits an effective bandwidth of 18 GHz (from 0.333 THz to 0.351 THz), with an average insertion loss of −1.3 dB and a reflection coefficient below −10 dB across the entire operating band. These results confirm the low-loss and well-matched guided-wave performance of the TW, establishing a solid foundation for the subsequent measurement of the TLWA.

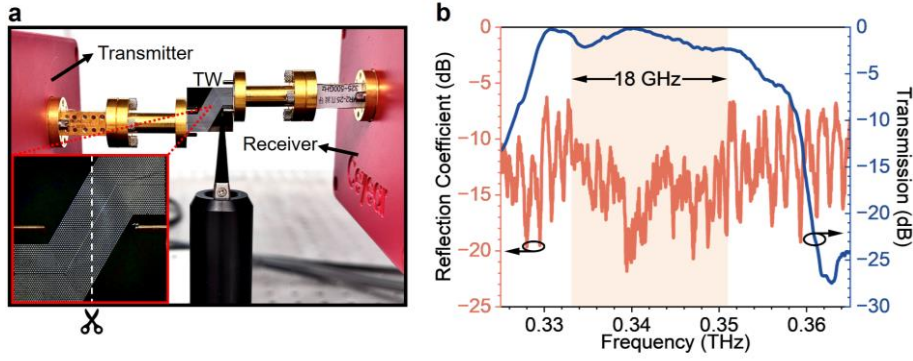

**FIGURE 4** | Presentation of the experimental setup and results of the proposed TW. (a) Presentation of the photograph of the TW and the equipment required for measuring the reflection coefficient and transmission spectrum. (b) Measured reflection coefficient and transmission spectrum of the TW.

The far-field normalized radiation pattern, reflection coefficient, and realized gain of the proposed TLWA are also measured using the VNA, with the experimental setup illustrated in Figure 5a. It is worth noting that the TLWA is intentionally designed with a larger structural footprint to maintain the best performance, while a smaller TLWA remains functional with slightly higher reflection coefficient. Owing to the polarization orientation of the waveguide connected to the frequency extender module, a twist waveguide is inserted between the waveguide and the antenna under test to ensure polarization matching. By rotating the receiving extender module via a mechanical rail, the far-field normalized radiation pattern is measured at $\theta = 90°$ over $\varphi$ ranging from $-90°$ to $90°$, as shown in Figure 5d. In addition, the reflection coefficient and the realized gain at $\theta = 90°$ and $\varphi = 0°$ are measured, with the results presented in Figure 5e.

To ensure accurate far-field normalized radiation pattern and realized gain measurements, we first determine the effective radiation region of the TLWA through simulation. Figure 5b depicts the magnetic field intensity distribution, where the bright regions represent 87% of the total energy confinement [30]. From this distribution, the equivalent radiation aperture dimensions in the *xoy*-plane ($D_1 = 5.25$ mm) and *yoz*-plane ($D_2 = 1.39$ mm) are extracted, as indicated in Figure 5b. Based on the derivation provided in Supporting Information S5, the following relation is obtained:

$$\begin{cases} \Delta G_{HA}(d) \neq \Delta S_{21,ref}(d) - \Delta S_{21,TLWA}(d), d \in \left(0.62\sqrt{D_1^3/\lambda}, 2D_1^2/\lambda\right] \\ \Delta G_{HA}(d) = \Delta S_{21,ref}(d) - \Delta S_{21,TLWA}(d), d \in \left(2D_1^2/\lambda, +\infty\right) \end{cases} \quad (2)$$

In equation (2), $\lambda$ is the wavelength of the TLWA, $G_{HA}(d)$ refers to the realized gain of the horn antenna at the radiation distance $d$, $S_{21,ref}(d)$ refers to the transmission gain received by the horn antenna when the horn antenna is transmitting at the radiation distance $d$, and $S_{21,TLWA}(d)$ refers to the transmission gain received by the horn antenna when the TLWA is transmitting at the radiation distance $d$. $\Delta$ represents the difference between the values measured at radiation distances $d$ and $d + \Delta d$. Using (2), the near-field to far-field transition distance of the TLWA is determined both theoretically and experimentally to be $d_1 = 62$ mm. Accordingly, in the far-field pattern measurement, the receiving extender module is placed at a distance greater than $d_1$. Since the receiving horn antenna itself has a near-field to far-field transition distance of $d_2 = 111$ mm and $d_2 > d_1$, the realized gain measurement is performed at a distance exceeding $d_2$ to guarantee far-field conditions. The realized gain was obtained using the comparison method, as detailed in Supporting Information S5. Figure 5c presents the calculated realized gain of the TLWA as a function of radiation distance, derived from the comparison method and equation (2). The results show that once the radiation distance exceeds $d_1$, the calculated realized gain stabilizes, which is consistent with the measured results.

Ultimately, the maximum realized gain reaches 12.5 dBi at $\theta = 90°$, and the 3-dB average beamwidth of the main lobe is 7°. The SLSR ($\theta = 80°$ and $100°$) of the TLWA exceeds 15 dB throughout the entire operating band (see Supporting Information S6 for details). The above results confirm the high-gain sharp single-beam characteristic of the TLWA. The reflection coefficient of the TLWA remains below –10 dB across the entire operating band, consistent with the well-matched condition obtained in the TW measurement. The TLWA also exhibits an effective bandwidth of 19 GHz (from 0.331 THz to 0.35 THz), with a fractional bandwidth substantially exceeding that of conventional terahertz leaky-wave antennas [37, 38]. This advantage stems from its unique low-dispersion property enabled by K-valley radiation. The main-lobe direction can be expressed by the following equation:

$$\varphi = \arcsin\left(\frac{K - k_{VPC}}{k_0}\right) \quad (3)$$

In equation (3), $k_{VPC}$ and $k_0$ represent the momentum module of the AB-type edge state and free space, respectively. Supporting Information S7 provides a detailed comparison between simulated and measured far-field normalized radiation patterns, demonstrating that the main-lobe direction of the TLWA varies only 2° across the entire operating band. Moreover, the simulated results show that the TLWA has an average total efficiency of 72%, as detailed in Supporting Information S8. These results confirm that the proposed TLWA features low dispersion, wide effective bandwidth, and high-efficiency single-beam characteristics, making it well suited for high-performance terahertz applications.

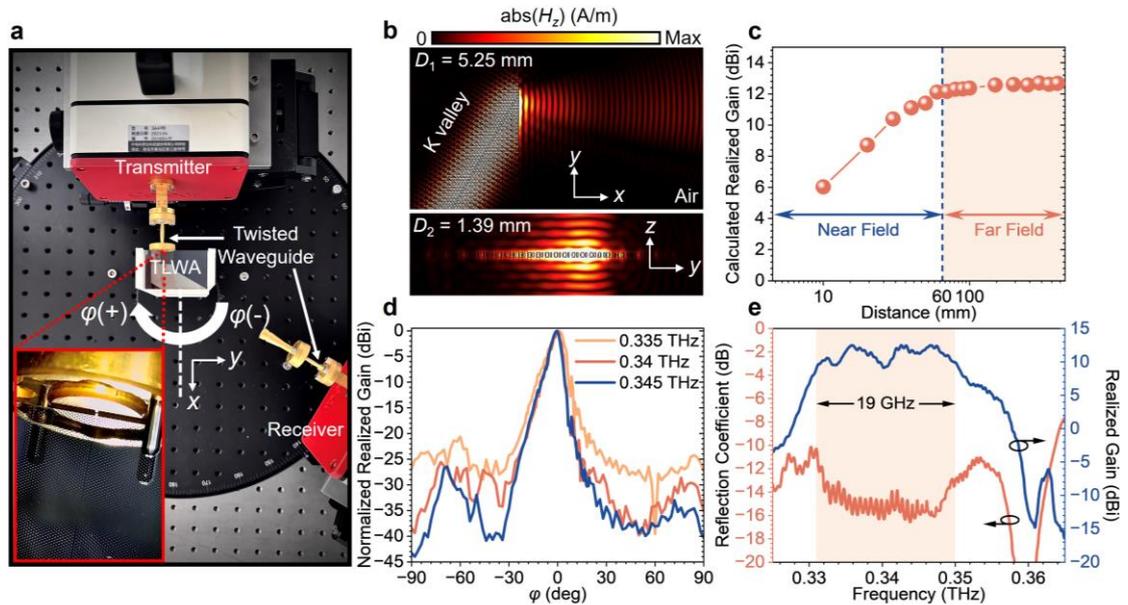

**FIGURE 5** | Presentation of the experimental setup and results of the proposed TLWA. (a) Presentation of the photograph of the TLWA and the equipment and methods required for measuring the reflection coefficient and far-

field parameters. (b) Simulated magnetic field distributions of the TLWA for calculating the maximum equivalent radiation aperture. (c) Calculated realized gains of different radiation distances of the TLWA. (d) Measured far-field normalized radiation patterns ($\theta = 90°$) of the TLWA at different frequencies. (e) Measured reflection coefficient and realized gain of the TLWA.

## 2.4 | Terahertz On-chip and Wireless Communication

The proposed TW exhibits low-loss and well-matched guided-wave performance. To establish a foundation for the subsequent terahertz communication demonstration of the proposed TLWA, we first characterize the terahertz communication performance of the TW. Figure 6a and 6b illustrate the framework of the terahertz communication system of the TW. At the transmitter side, a data stream is modulated using 16 quadrature amplitude modulation (16QAM) at a data rate of 60 Gbps. Following the procedure detailed in the Experimental Section, the terahertz carrier obtained by the uni-traveling-carrier photodiode (UTC-PD) is coupled into the TW via a rectangular metal waveguide and a taper coupler, propagates through the TW, and is coupled out by another identical taper coupler to obtain the received signal. The received signal is down-converted and subsequently demodulated. The demodulation results are presented in Figure 6c and Figures 6d$_1$–d$_2$. It can be seen that as the optical power injected into the UTC-PD increases from 5 dBm to 9 dBm, the constellation diagrams progressively sharpen, and the bit error rate (BER) decreases from $3.25\times10^{-2}$ to $3.1\times10^{-3}$. When the optical power exceeds 6 dBm, the BER falls below soft-decision forward-error-correction (SD-FEC) threshold, thereby enabling error-free transmission [39]. These results demonstrate the excellent terahertz communication capability of the proposed TW.

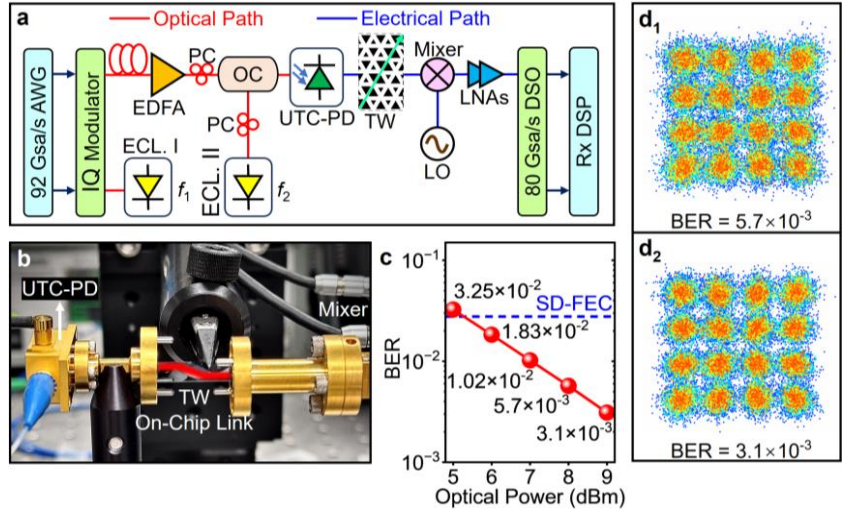

**FIGURE 6** | Terahertz communication system framework of the proposed TW and the relevant measured results of modulation and demodulation. (a) The terahertz communication system framework of the TW with optical transmitter and electrical receiver. (b) Detailed photograph of the on-chip terahertz communication link. (c) BER with different optical power under 60 Gbps 16QAM modulation format. (d$_1$-d$_2$) Constellation diagrams under 60 Gbps 16QAM modulation format at optical power of 8 dBm and 9 dBm.

The proposed TLWA exhibits a sharp single-beam radiation pattern, which effectively suppresses interference from undesired directions ($\varphi \neq 0°$) and ensures the quality of the received signal. Based on this characteristic, we employ the TLWA as a receiving antenna. A commercial horn antenna, with its highly concentrated beam in the $\theta = 90°$ direction, is used as the transmitter to compensate for the energy loss caused by the absence of terahertz low-noise amplifiers and collimating lenses in the communication link. Figures 7a and 7b illustrate the framework of the terahertz communication system of the TLWA. At the transmitter side, a data stream is modulated using quadrature phase shift keying (QPSK) at a data rate of 24 Gbps. Following the procedure detailed in the Experimental Section, the terahertz carrier obtained by the UTC-PD is radiated by the horn antenna, propagates through 2.5 mm free space, and is received by the TLWA for down-conversion and demodulation. The demodulation results are presented in Figure 7c and Figures $7d_1$–$d_2$. It can be seen that as the optical power injected into the UTC-PD increases from 5 dBm to 9 dBm, the constellation diagrams progressively sharpen, and the BER decreases from $1.122×10^{-1}$ to $9.8×10^{-3}$. When the optical power exceeds 8 dBm, the BER falls below SD-FEC threshold, thereby enabling error-free transmission.

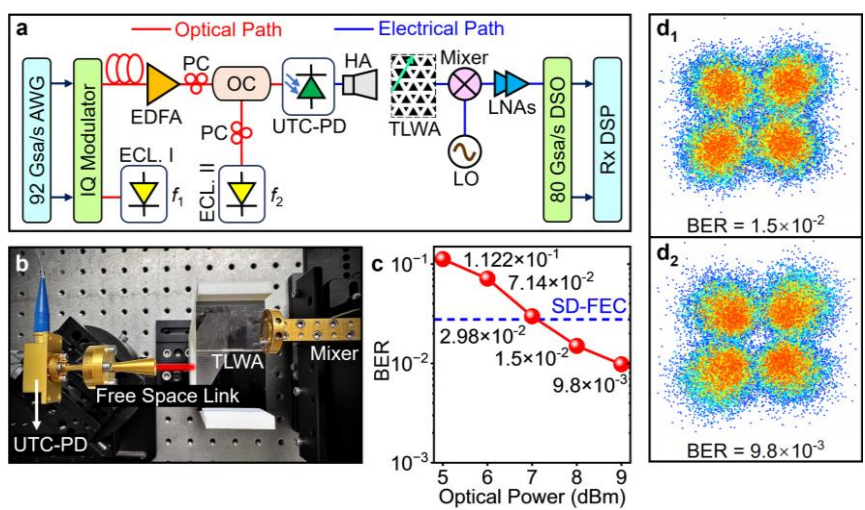

**FIGURE 7** | Terahertz communication system framework of the proposed TLWA and the relevant measured results of modulation and demodulation. (a) The terahertz communication system framework of the TLWA with optical transmitter and electrical receiver. (b) Detailed photograph of the wireless terahertz communication link. (c) BER with different optical power under 24 Gbps QPSK modulation format. ($d_1$-$d_2$) Constellation diagrams under 24 Gbps QPSK modulation format at optical power of 8 dBm and 9 dBm.

To further demonstrate the advantage of using the proposed TLWA as a receiving antenna, a comparative experiment is conducted in Supporting Information S9. The transmitting horn antenna and the TLWA are replaced by two rectangular metal waveguides under identical conditions. In this case, no signal could be demodulated. Supporting Information S10 presents transmission spectra at different radiation distances of the TLWA and compares the performance of the TLWA with other terahertz antennas. These results further underscore the unique value of the proposed TLWA as a receiving antenna for high-speed terahertz communication. Thus far, the proposed TW and TLWA have been comprehensively characterized. The results demonstrate that whether configured as a TW or as a TLWA, the proposed device exhibits significant application potential in the terahertz band. This bifunctionality provides a valuable foundation for the design of highly integrated and modular terahertz on-chip systems.

# 3 | Conclusions

We have proposed and experimentally demonstrated a topological valley-reshaped device that inherently possesses bifunctionality: derived from a complete TW via truncation at a specific angle, the proposed device inherently features guided-wave function serving for the TW while independently functioning as a TLWA without any additional structures. The TW exhibits low-loss guided-wave performance with an 18 GHz operating bandwidth and supports error-free transmission up to 60 Gbps. By truncating the same structure along specific directions, the valley-locked edge state naturally transitions into a leaky mode, achieving momentum-matching with the free-space light line. Through valley-reshaping of the opposite edge state, the TLWA achieves simultaneous suppression of far-field side lobes and reduction of reflection, yielding a clean single-beam radiation pattern with 15 dB SLSR, a peak gain of 12.5 dBi, and a 19 GHz operating bandwidth. Owing to its low-dispersion property enabled by K-valley radiation, the main-lobe direction varies by only 2° across the entire operating band, enabling error-free free-space transmission at 24 Gbps as a receiver. These results demonstrate that the proposed device successfully bridges the long-standing gap between high-performance on-chip waveguiding and efficient free-space radiation within the same topological platform. The inherent reconfigurability between guided-wave and radiative functionalities enabled by valley-reshaping provides a valuable foundation for highly integrated and versatile terahertz on-chip systems, simultaneously paving the way for advanced applications in 6G communications and flexible photonic circuits.

# 4 | Experimental Section

## 4.1 | Experimental Setup for Terahertz Communication System

At the transmitter side, a baseband data stream is generated by a 90 GSa/s arbitrary waveform generator (AWG) and modulated onto the optical carrier from an external cavity laser (ECL. I) via an IQ modulator. The center frequency of the optical carrier from ECL. I is 193.548 THz. The modulated optical signal is then amplified by an erbium-doped fiber amplifier (EDFA) and combined with an optical monophonic signal from another external cavity laser (ECL. II) through an optical coupler (OC). The frequency of the optical monophonic signal from ECL. II is 193.21 THz. The combined optical signals are fed into UTC-PD, where nonlinear difference-frequency generation produces a terahertz carrier centered at 0.338 THz (see Supporting Information S9 for optical spectrum before injecting into the UTC-PD). The generated terahertz carrier is guided through a WR-3.4 waveguide coupled to the UTC-PD for propagation. At the receiver side, the carrier enters a mixer through a WR-2.8 waveguide, where it is down-converted to an intermediate frequency of 18 GHz using a local oscillator (LO) signal at 0.32 THz (generated via frequency multiplication). The down-converted signal is amplified by cascading coaxial intermediate frequency low-noise amplifiers (LNAs) and passed through attenuators before being captured by an 80 GSa/s digital storage oscilloscope (DSO). Finally, the captured data are processed offline using digital signal processing (DSP) for demodulation and BER analysis.

## 4.2 | Devices Fabrication

Devices were fabricated on 200-μm-thick high-resistivity (>10,000 Ω·cm) silicon wafers. After standard cleaning, an aluminum layer was deposited on the backside by magnetron sputtering. The wafer was then anodically bonded to a glass substrate (380 °C, 10,000 N, 1.2 h). Photolithography defined the photonic structures using AZ4620 photoresist. Deep reactive ion etching (Bosch process) transferred the pattern into silicon with alternating etch ($SF_6$, 2.2 s) and passivation ($C_4F_8$, 2.0 s) cycles. The resist and backside aluminum were subsequently removed, yielding the final devices. The fabrication results of the devices are provided in Supporting Information S11.

**Authors' Contributions**

S. H. conceived the idea. Y. W. developed the theoretical framework, designed the device, and performed the numerical simulations under guidance of S. H. Y. W. and F. C. led the device characterization. Y. W. and Z. W. conducted the THz communication experiments. H. S. set up the experimental apparatus. B. G., and S. L., helped the fabrication. J. T. supervised the THz communication experiments. S. H. and Y. W. wrote the manuscript. S. H. and H. C. supervised the overall project and provided feedback on the manuscript. All authors discussed the results and contributed to the final manuscript.

**Acknowledgements**

S. H. acknowledges the National Key R&D Program of China (2024YFB2808200), the National Natural Science Foundation of China (62475230), and the Excellent Young Scientists Fund Program (Overseas) of China. J. T. acknowledges the National Natural Science Foundation of China (62275208). All authors acknowledge the Nanjing Nuozhijie Electronic Technology Co., Ltd.

**Conflicts of Interest**

The authors declare no conflicts of interest.

**Data Availability Statement**

The data that support the findings of this study are available from the corresponding author upon reasonable request.

**References**

1. H. Zhang, X. Huang, X. Guo, et al., "Terahertz Sensing, Communication, and Networking: A Survey," *IEEE Transactions on Network Science and Engineering* 13 (2026): 501-521.

2. W. Li, H. Zeng, L. Huang, et al., "A Review of Terahertz Solid-State Electronic/Optoelectronic Devices and Communication Systems," *Chinese Journal of Electronics* 34, no. 1 (2025): 26-48.

3. W. Jiang, Q. Zhou, J. He, et al., "Terahertz Communications and Sensing for 6G and Beyond: A Comprehensive Review," *IEEE Communications Surveys & Tutorials* 26, no. 4 (2024): 2326-2381.

4. I. F. Akyildiz, C. Han, Z. Hu, S. Nie, and J. M. Jornet, "Terahertz Band Communication: An Old Problem Revisited and Research Directions for the Next Decade," *IEEE Transactions on Communications* 70, no. 6 (2022): 4250-4285.

5. G. Liu, B. Zhang, H. Zhu, and Y. Zhang, "An Isolated H-Plane Terahertz Waveguide T-Junction With Corrugated Microstrip Loads," *IEEE Transactions on Terahertz Science and Technology* 15, no. 1 (2025): 133-136.

6. M. Yamazaki, Y. Sugimoto, K. Sakakibara, and N. Kikuma, "Broadband Differential-Line-to-Waveguide

Transition in Multi-Layer Dielectric Substrates With an X-Shaped Patch Element in 280 GHz Band," *IEEE Transactions on Microwave Theory and Techniques* 71, no. 6 (2023): 2616-2624.

7. B. Yuan, P. Wu, Z. Yu, and C. Hao, "Wideband End-Wall Transition From Microstrip to Waveguide With via-Less Choke Structure for Terahertz Application," *IEEE Transactions on Terahertz Science and Technology* 12, no. 3 (2022): 317-320.

8. X. Zhang, S. Li, C. Yu, W. Gao, and M. L. N. Chen, "Reconfigurable Topological Power Router Based on Interferometric Edge States," *IEEE Transactions on Antennas and Propagation* (2026).

9. R. Zhou, X. Shi, H. Lin, et al., "Super-Robust Telecommunications Enabled by Topological Half-Supermodes," *Advanced Science* 13, no. 13 (2026): e15157.

10. H. Wang, H. Shi, W. E. I. Sha, et al., "Design of High-Isolation Topological Duplexer Utilizing Dual-Edge State Topological Waveguides," *IEEE Transactions on Antennas and Propagation* 72, no. 11 (2024): 8802-8809.

11. R. Banerjee, A. Kumar, T. C. Tan, et al., "On- chip amorphous terahertz topological photonic interconnects," *Science Advances* 11, no. 25 (2025): eadu2526.

12. Z.-H. Xu, J. He, X. Hao, et al., "Quantitative Terahertz Communication Evaluation of Compact Valley Topological Photonic Crystal Waveguides," *ACS Photonics* 12, no. 4 (2025): 1822-1828.

13. Y. Yang, Y. Yamagami, X, Yu, et al., "Terahertz topological photonics for on-chip communication," *Nature Photonics* 14 (2020): 446-451.

14. R. Jia, W. Wang, Y. J. Tan, et al., "Tunable Topological Directional Supercoupler and Applications in THz On-Chip Communication," *Laser & Photonics Reviews* 20, no. 1 (2026): e01209.

15. W.-Y. Wang, H. Ren, Z.-H. Xu, H. Chen, Y. Li, and S. Xu, "Integrated terahertz topological valley-locked power divider with arbitrary power ratios," *Optics Letters* 49, no. 19 (2024): 5579-5582.

16. H. Wang, G. Tang, Y. He, et al., "Ultracompact topological photonic switch based on valley-vortex-enhanced high-efficiency phase shift," *Light: Science & Applications* 11 (2022): 292.

17. X. Liu, J. Huang, H. Chen, et al., "Terahertz topological photonic waveguide switch for on-chip communication," *Photonics Research* 10, no. 4 (2022): 1090-1096.

18. M. Gupta, A. Kumar, P. Pitchappa, et al., "150 Gbps THz Chipscale Topological Photonic Diplexer," *Advanced Materials* 36 (2024): 2309497.

19. H. Ren, S. Xu, Z. Lyu, et al., "Terahertz flexible multiplexing chip enabled by synthetic topological phase transitions," *National Science Review* 11, no. 8 (2024): nwae116.

20. A. Kumar, M. Gupta, P. Pitchappa, et al., "Phototunable chip-scale topological photonics: 160 Gbps waveguide and demultiplexer for THz 6G communication," *Nature Communications* 13 (2022): 5404.

21. W. Wang, Y. J. Tan, T. C. Tan, et al., "On-chip topological beamformer for multi-link terahertz 6G to XG wireless," *Nature* 632 (2024): 522-527.

22. K. Chen, W. Wang, Y. J. Tan, Z. Shen, R. Jia, and R. Singh, "Ultralow-Backscattering Induced Mode Splitting in Chipscale Valley Photonic Topological Cavities," *Laser & Photonics Reviews* 20, no. 1 (2026): e03189.

23. W. Wang, Z. Shen, Y. J. Tan, K. Chen, and R. Singh, "On-chip topological edge state cavities," *Light: Science &*

*Applications* 14 (2025): 330.

24. Z. Shen, Y. J. Tan, W. Wang, et al., "Interface Topology Driven Loss Minimization in Integrated Photonics: THz Ultrahigh-Q Cavities and Waveguides," *Advanced Materials* 37, no. 35 (2025): 2503460.

25. W. Wang, Y. J. Tan, P. Szriftgiser, G. Ducournau, and R. Singh, "On-chip topological leaky-wave antenna for full-space terahertz wireless connectivity," *Nature Photonics* 20 (2026): 317-323.

26. H. Wang, H. Shi, W. E. I. Sha, et al., "Terahertz high-gain leaky-wave antenna utilizing valley topological photonic crystals with line defects," *Journal of Physics D: Applied Physics* 58 (2025): 215104.

27. S. Kumar, M. Gupta, T. C. Tan, A. Alphones, and R. Singh, "Dual-band Topological Filter Antenna on a Silicon Chip for Terahertz wireless Communication," *IEEE Transactions on Terahertz Science and Technology* (2026).

28. S. Kumar, R. Jia, Y. J. Tan, et al., "Topological Berry Antenna on a Silicon Chip for Terahertz Wireless Communication," *Advanced Photonics Research* 6 (2025): 2500123.

29. R. Jia, S. Kumar, T. C. Tan, et al., "Valley-conserved topological integrated antenna for 100-Gbps THz 6G wireless," *Science Advances* 9, no. 44 (2023): eadi8500.

30. Z. Zhang, D. Liao, Y. Li, et al., "Energy-Efficient Integrated Photonic Topological Devices," *Laser & Photonics Reviews* 19, no. 1 (2025): 2400567.

31. F. Gao, H. Xue, Z. Yang, et al., "Topologically protected refraction of robust kink states in valley photonic crystals," *Nature Physics* 14 (2018): 140-144.

32. S. Yuan, Z. Z. Ding, X. G. Zhang, D. J. Wang, J. K. Sun, and W. X. Jiang, "Laser-Tracking-Modulated Microwave Temporal Metasurfaces for Mobile Hybrid Wireless Communications," *Progress In Electromagnetics Research* 184 (2025): 24-31.

33. H. Qiu, L. Fang, R. Xi, et al., "Wideband High Gain Lens Antenna Based on DeepLearning Assisted Near-Zero Refractive Index Metamaterial," *Progress In Electromagnetics Research* 182 (2025): 13-25.

34. H. Chen, H. Ren, W. Wang, et al., "Terahertz chiral edge states enable inner-chip state transition and interchip communications over wireless terminals," *Chinese Optics Letters* 22, no. 10 (2024): 103701.

35. L. Cheng, X. Y. Li, J. L. Su, et al., "Variational Quantum Algorithm for Photonic Crystals," *Progress In Electromagnetics Research* 184 (2025): 1-13.

36. L. He, D. Liu, H. Zhang, et al., "Topologically Protected Quantum Logic Gates with Valley-Hall Photonic Crystals," *Advanced Materials* 36 (2024): 2311611.

37. Y. Liu, W. Chen, J. Huang, M. Zhang, H. Su, and H. Liang, "A High-Gain Terahertz Leaky-Wave Meta-Antenna Based on Parabolic Reflector Feeding," *IEEE Antennas and Wireless Propagation Letters* 24, no. 12 (2025): 4975-4979.

38. H. Guerboukha, M. Sakaki, R. Shrestha, et al., "3D-Printed Photonic Crystal Sub-Terahertz Leaky-Wave Antenna," *Advanced Materials Technologies* 9, no. 6 (2024): 2300698.

39. D. Chang, F. Yu, Z. Xiao, et al., "LDPC Convolutional Codes using Layered Decoding Algorithm for High Speed Coherent Optical Transmission," *OFC/NFOEC, IEEE*, Los Angeles, CA, USA (2012).

# Topological Valley-Reshaped Device: Bifunctional Waveguiding and Single-Beam Leaky-Wave Radiation for Terahertz Communication

Yulun Wu[1,2,3] | Ziwei Wang[4,5] | Faqian Chong[1,2,3] | Hua Shao[1,2,3] | Bingtao Gao[1,2,3] | Shilong Li[1,2,3] | Jin Tao[4,5] | Hongsheng Chen[1,2,3,6] | Song Han[1,2,3,7]

[1]Innovative Institute of Electromagnetic Information and Electronic Integration, College of Information Science & Electronic Engineering, Zhejiang University, Hangzhou, China | [2]State Key Laboratory of Extreme Photonics and Instrumentation, ZJU-Hangzhou Global Scientific and Technological Innovation Center, Zhejiang University, Hangzhou, China | [3]International Joint Innovation Center, The Electromagnetics Academy at Zhejiang University, Zhejiang University, Haining, China | [4]State Key Laboratory of Optical Communication Technologies and Networks (OCTN), Wuhan, China | [5]Information Communication Technologies Group Corporation (CICT), Wuhan, China | [6]Key Lab. of Advanced Micro/Nano Electronic Devices & Smart Systems of Zhejiang, Jinhua Institute of Zhejiang University, Zhejiang University, Jinhua, China | [7]Zhejiang Key Laboratory of Advanced Micro-nano Transducers Technology, College of Information Science and Electronic Engineering, Zhejiang University, Hangzhou, China

**Correspondence:** Hongsheng Chen (hansomchen@zju.edu.cn) | Song Han (song.han@zju.edu.cn)

Yulun Wu and Ziwei Wang contributed equally to this work.

## S1 | Derivation of the Radiation Principle of the Proposed Device

In this section, we focus on discussing how the proposed device achieves radiation through momentum matching. The refractive index of the air can be approximately considered as $n_0 = 1$. The effective refractive index of the K valley can be expressed as [S1]:

$$n = \frac{2c}{3an_0 f} \quad (S1)$$

In (S1), $c$ represents the speed of light, $a$ represents the lattice constant of the device and $f$ represents the leaky-wave frequency of the device. According to (S1), $n = 2.5 > n_0$. Thus, the momentum vectors in real space and reciprocal space are illustrated in Figure S1. The momentum matching condition at the interface between the AB-type edge state (mentioned in the main text) and air can be expressed as [S1]:

$$\boldsymbol{k}_{vpc} \cdot \boldsymbol{e}_t = \boldsymbol{k}_{air} \cdot \boldsymbol{e}_t \quad (S2)$$

In (S2), $\boldsymbol{k}_{VPC}$ and $\boldsymbol{k}_{air}$ represent the momentum of the AB-type edge state and free space, respectively. And $e_t$ represents the truncation direction along the device-air interface. Since both sides of equation (S2) are 0 regardless of the magnitude mismatch, leaky-wave radiation is enabled.

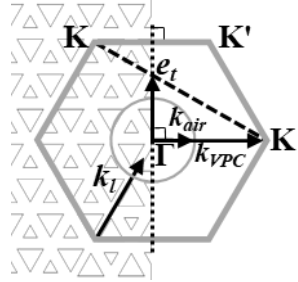

**FIGURE S1** | Momentum distributions of the real space and reciprocal space. The position of the proposed device corresponds to the directions of momentums, but the relative scale between real and reciprocal space is not preserved.

## S2 | Basics of Topological Valley Photonics and Simulated Results Corresponding to Different $\delta$ Values

The proposed device consists of a rhombic unit cell with lattice constant of a = 242.5 μm. Each unit cell contains two mirror-symmetric equilateral triangular air holes with side lengths $l_1$ and $l_2$, respectively. We define $\delta = l_1 - l_2$. As shown in Figure S2a, when $\delta = 0$, the lattice preserves inversion symmetry with $C_{6v}$ symmetry, yielding a pair of degenerate Dirac points at the high-symmetry points K and K' in the Brillouin zone. When $\delta \neq 0$, inversion symmetry is broken, opening a topological bandgap at the Dirac points. In general, a larger $\delta$ leads to a wider topological bandgap.

Figure S2b shows the global design of the proposed device. The device is composed of two sections. The guided-wave section (dark blue line) is formed by countless defects composed of two small mirror-symmetric equilateral triangular air holes, and it supports BA-type edge state. The leaky-wave section (dark orange line) is formed by countless defects composed of two large mirror-symmetric equilateral triangular air holes, and it supports AB-type edge state. To visually demonstrate the distinct parities of the above two valley-locked edge states, Figure S2b presents their simulated magnetic field distributions. In operation, the device is excited by $TE_{10}$ mode propagating from a rectangular metal waveguide via a silicon taper coupler. Two identical devices can form a complete topological waveguide (TW) by rotationally symmetric placing, while a single device independently functions as a topological leaky-wave antenna (TLWA).

We refer the mode profiles of the upper and lower bands at the K and K' valleys as type A and type B, respectively, as illustrated in Figure S2b. These two mode profiles have opposite phase distributions, reflecting their intrinsic symmetry within the valley-locked system. In conventional designs, topological waveguides or antennas that support two valley-locked edge states typically adopt the same $\delta$ value for both type A and type B lattices. However, to enhance the effective bandwidth of the device (defined as the overlap between the impedance bandwidth and the transmission or radiation bandwidth), we intentionally assign different $\delta$ values to the lattices supporting the AB-type ($\delta = 0.3a$) and BA-type ($\delta = 0.25a$) edge states. As shown in Figure S2c, increasing the side length $l_2$ of the small triangular holes (i.e., increasing $\delta$) has a negligible effect on the propagation frequency of the AB-type edge state, but significantly raises the high-frequency cutoff of the BA-type edge state. Consequently, the impedance operating band shifts toward the transmission operating band, effectively enhancing the achievable effective bandwidth.

Figure S3 shows the simulated reflection coefficients, transmissions, and realized gains corresponding to different $\delta$ values of the proposed topological waveguide (TW) and the proposed topological leaky-wave antenna (TLWA). The effective bandwidth is maximized when $\delta = 0.3a$ for the AB-type edge state and $\delta = 0.25a$ for the BA-type edge state.

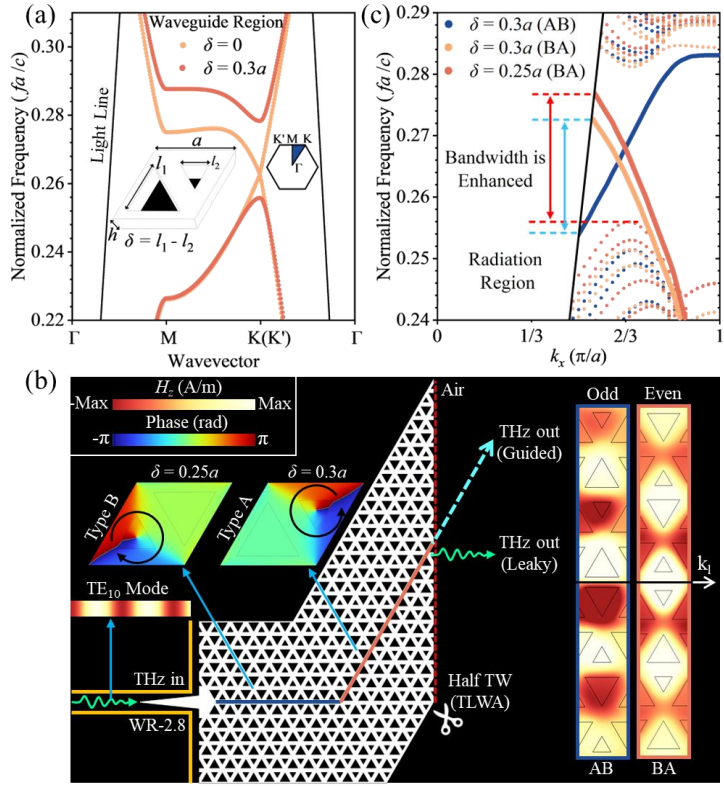

**FIGURE S2** | Global design and partial theory explanation of the proposed device, the device can be regarded as either the proposed TLWA (as a single unit) or half of the proposed TW (when paired with an identical device). (a) Band diagrams of the first Brillouin zone under different $\delta$ and structural parameters: $a = 242.5$ μm, $h = 200$ μm, $l_1 = 0.65a$, $\delta = l_1 - l_2$. (b) The top view of the device and the magnetic field and phase distributions. (c) Band diagrams of AB-type edge state and BA-type edge state under different $\delta$.

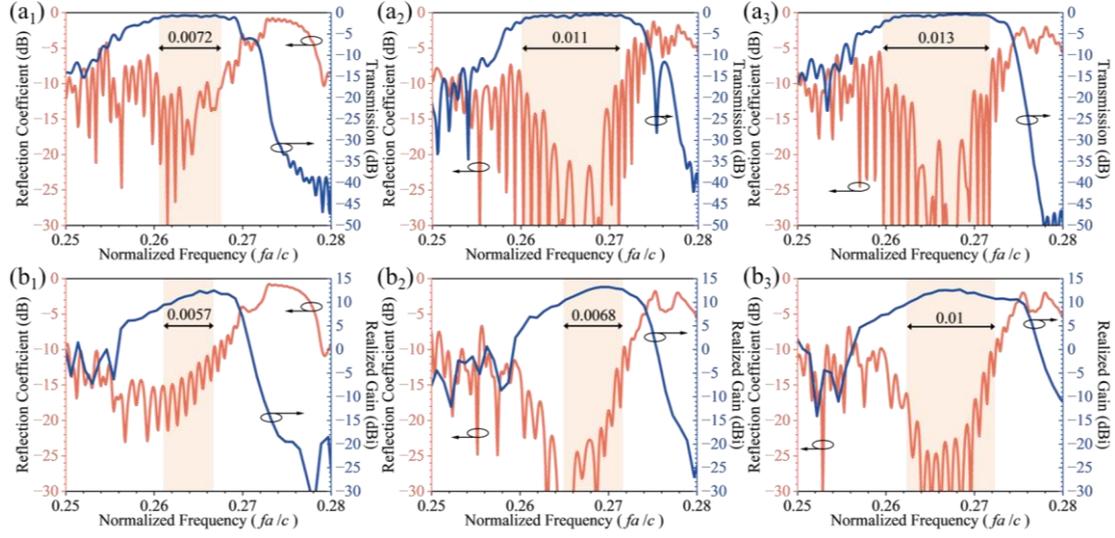

**FIGURE S3** | Simulated reflection coefficients and realized gains corresponding to different $\delta$ values. ($a_1$) Simulated results of AB-type ($\delta = 0.3a$) and BA-type ($\delta = 0.3a$) edge states of the TW. ($a_2$) Simulated results of AB-type ($\delta = 0.25a$) and BA-type ($\delta = 0.25a$) edge states of the TW. ($a_3$) Simulated results of AB-type ($\delta = 0.3a$) and BA-type ($\delta = 0.25a$) edge states of the TW. ($b_1$) Simulated results of AB-type ($\delta = 0.3a$) and BA-type ($\delta = 0.3a$) edge states of the TLWA. ($b_2$) Simulated results of AB-type ($\delta = 0.25a$) and BA-type ($\delta = 0.25a$) edge states of the TLWA. ($b_3$) Simulated results of AB-type ($\delta = 0.3a$) and BA-type ($\delta = 0.25a$) edge states of the TLWA.

## S3: Derivation of Equation (1) and More Information About Phase Gradients

According to Helmholtz equation, we can derive Kirchhoff's far-field diffraction formula using Green's function method:

$$U(P) = \frac{1}{4\pi} \iint_S \left[ U(Q) \frac{\partial}{\partial n}\left(\frac{e^{jkr}}{r}\right) - \frac{e^{jkr}}{r} \frac{\partial U(Q)}{\partial n} \right] dS \tag{S3}$$

In (S3), $S$ is the radiation aperture plane for integral, $n$ is the normal vector perpendicular to this plane, and $k$ is the momentum in transmission medium. Since the radiation aperture of the proposed device is mainly distributed on the $xoy$ plane, it is assumed that the coordinates of point $Q$ on the aperture are $(x, y, 0)$, and the coordinates of the observation point $P$ are $(x_P, y_P, z_P)$. $U(P)$ and $U(Q)$ represent the field intensities of $P$ and $Q$, respectively. Then, we let:

$$\begin{cases} r = \sqrt{(x_p - x)^2 + (y_p - y)^2 + z_p^2} \\ r_0 = \sqrt{x_p^2 + y_p^2 + z_p^2} \end{cases} \tag{S4}$$

By making amplitude approximation and phase approximation for $r$, the following relationship can be obtained:

$$r = r_0 - \sin\theta\cos\varphi\, x - \sin\theta\sin\varphi\, y \tag{S5}$$

Substitute (S5) into (S3), and we obtain the Fraunhofer's far-field diffraction formula:

$$U(P) = \frac{e^{jkr_0}}{j\lambda r_0} \iint_S U(x,y) e^{-jk(\sin\theta\cos\varphi x + \sin\theta\sin\varphi y)} dxdy \tag{S6}$$

Applying (S6) to the equivalent one-dimensional line source of the radiation aperture of the device, the integration range is $D_1$. Therefore, the far-field intensity of magnetic field $H(\theta, \varphi)$ can be expressed by the following equation:

$$H(\theta,\varphi) = \frac{Ce^{jk_0 r_0}}{r_0} \int_{-D_1/2}^{D_1/2} H_z(y) e^{-jk_0 y \sin\theta \sin\varphi} dy \tag{S7}$$

In (S7), $k_0$ is the momentum in free-space and $C$ is constant. Since $H_z(y)$ may be composed of the superposition of magnetic fields of different modes corresponding to multiple phase gradients, each mode having an amplitude $A_i$ and a phase gradient $\beta_{y,i}$, and the radiation direction of the device forms a 60° angle with the guided-wave direction. Therefore, $H_z(y)$ can be written as:

$$H_z(y) = \sum_{i=1}^{n} A_i e^{\frac{\sqrt{3}}{2} j \beta_{l,i} y} \tag{S8}$$

Substitute (S8) into (S7), and we can obtain:

$$H(\theta,\varphi) = \frac{D_1 e^{jk_0 r_0}}{r_0} \sum_{i=1}^{n} B_i \operatorname{sinc}\left(\frac{\sqrt{3}D_1}{4}\beta_{l,i} - \frac{D_1}{2} k_0 \sin\theta \sin\varphi\right) \tag{S9}$$

In (S9), $\operatorname{sinc}(x) = \sin(x)/x$, $B_i$ is the weight of each phase gradient. Finally, we can obtain the normalized far-field intensity proposed in main text. Figure S4a shows the results of the calculated normalized radiation pattern components as $\beta_l$ varies. It can be seen that as $\beta_l$ increases from 0, the main lobe splits into two at $\beta_l = 0.2$. As $\beta_l$ continues to increase, the side lobe gradually moves to the negative $\varphi$ direction until the side lobes disappear at $\beta_l = 0.53$. Figure S4b$_1$-b$_3$ further validate the reasonableness of using Equation (1) to approximately explain the effect of phase gradient on far field by comparing the calculated and simulated results.

Figure S4c$_1$-c$_3$ shows simulated near-field phase and phase gradient distributions of BA-type valley-locked edge state, BA-AB-BA valley-locked edge state coupling, and BA-AB valley-locked edge state coupling, respectively. Combining the results of Figure 4e$_1$-e$_3$ in the main text, it can be seen that, no matter how many times the valley-locked edge state switches during the waveguiding process, the final $\beta_l$ distribution depends only on the first valley-locked edge state. This indicates that the valley-locked edge state guided-wave system is not a simple linear time-invariant (LTI) system, which may stem from its unique valley memory effect (or valley-dependent history). Therefore, the far-field radiation pattern can only be reshaped when the first valley-locked edge state is BA-type valley-locked edge state.

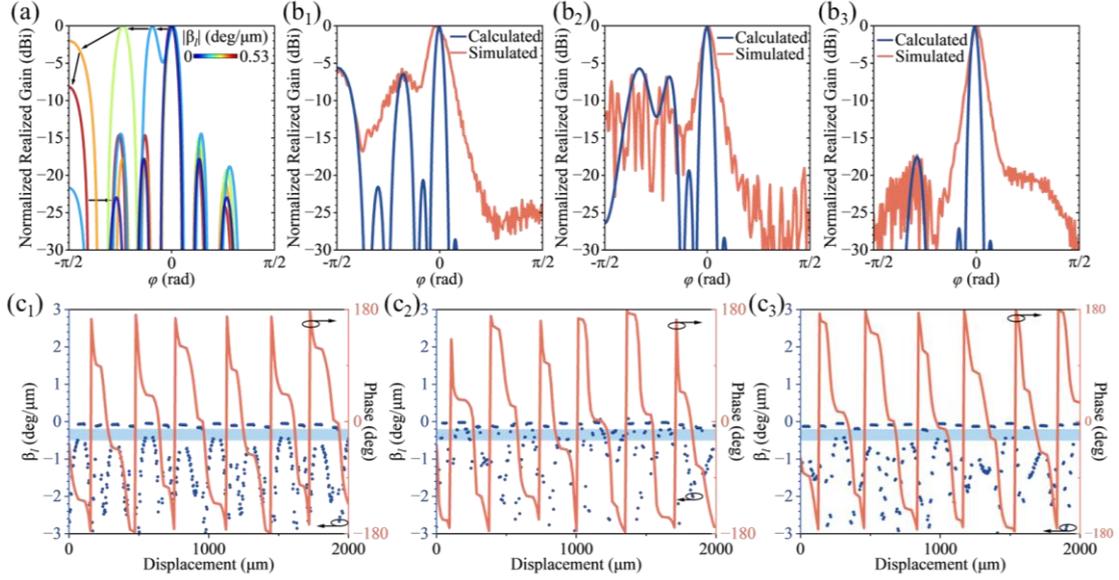

**FIGURE S4** | Calculated and simulated results corresponding to different combination effects of different valley-locked edge states. (a) Results of the calculated normalized radiation pattern components as $β_l$ varies. ($b_1$-$b_3$) Calculated and simulated normalized far-field radiation patterns with AB-type valley-locked edge state configuration, AB-BA-AB valley-locked edge state coupled configuration, and BA-AB valley-locked edge state coupled configuration, respectively. ($c_1$-$c_3$) Simulated near-field phase and phase gradient distributions of BA-type valley-locked edge state, BA-AB-BA valley-locked edge state coupling, and BA-AB valley-locked edge state coupling, respectively.

## S4 | Simulated Results of the Proposed Modified Dual-Beam TLWA

Based on the symmetry characteristics of the valley-locked edge state, by truncating the AB-type straight waveguide along both the 60° and –60° directions and incorporating valley-reshaping, it is possible to achieve radiation in the direction symmetrical to the axis of the topological defect structure for the original radiation direction. Therefore, dual-beam TLWA can be realized by modifying the proposed device, as shown in Figure S5. It can be seen that the modified TLWA can achieve dual-beam radiation without parasitic side lobes, while also having a reflection power as low as 1% and a large effective bandwidth at $φ = -60°$ and $φ = 60°$ simultaneously.

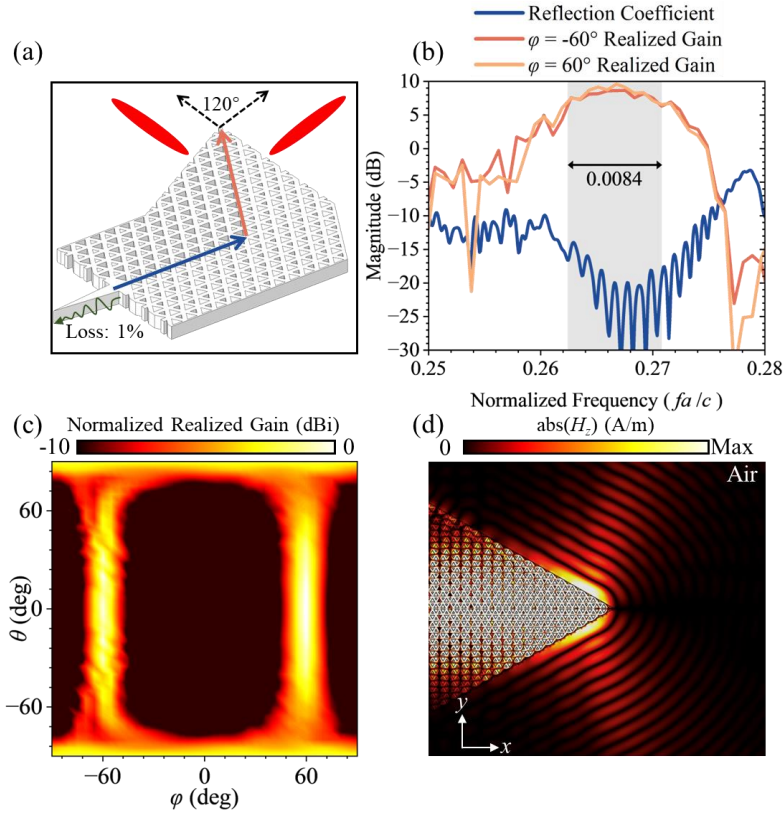

**FIGURE S5** | Simulated results of the proposed modified dual-beam TLWA. (a) The structure of the modified dual-beam TLWA. (b) Simulated reflection coefficient and realized gains at $\varphi = -60°$ and $\varphi = 60°$ of the modified dual-beam TLWA. (c) Simulated far-field normalized radiation pattern of the modified dual-beam TLWA. (d) Simulated magnetic field distributions of the modified dual-beam TLWA.

## S5 | Calculations of the Near Field and Far Field

Whether it is a horn antenna or the proposed TLWA, as long as it is an aperture antenna, the induction-field to near-field transition distance $d_{i-n}$ and the near-field to far-field distance $d_{n-f}$ are both satisfied by the following equation [S2]:

$$\begin{cases} d_{i-n} = 0.62\sqrt{D_{max}^3/\lambda} \\ d_{n-f} = 2D_{max}^2/\lambda \end{cases} \tag{S10}$$

In equation (S10), $\lambda$ is the wavelength, and $D_{max}$ represents the maximum radiation aperture dimension of the antenna. The TLWA satisfies $D_{max} = D_1 = 5.25$ mm (mentioned in the main text). Since the horn antenna is conical in shape, its aperture remains $D_{max} = 2R_0 = 7$ mm in diameter. Therefore, $d_{n-f}$ of the TLWA and the horn antenna can be calculated, with the results being $d_1 = 62$ mm and $d_2 = 111$ mm, respectively.

Figure S6 shows how the realized gain of the TLWA is measured. According to the Friis formula, the terahertz link in Figure S6 can be written as a pair of mathematical expressions:

$$\begin{cases} P_{Tx,TLWA} + G_{TLWA} - P_{loss} + G_{HA} = P_{Rx,TLWA} \\ P_{Tx,ref} + G_{HA} - P_{loss} + G_{HA} = P_{Rx,ref} \end{cases} \tag{S11}$$

In (S11), $P_{Tx,TLWA}$ and $P_{Rx,TLWA}$ represent the transmission power and the received power in the scenario of Figure S6a, and $P_{Tx,ref}$ and $P_{Rx,ref}$ represent the transmission power and the received power in the scenario of Figure S6b. $G_{TLWA}$ represents the realized gain of the TLWA and $G_{HA}$ represents the realized gain of the horn antenna. $P_{loss}$ represents the loss of the free-space link. To simplify the expression, we define:

$$\begin{cases} S_{21,TLWA} = P_{Rx,TLWA} - P_{Tx,TLWA} \\ S_{21,ref} = P_{Rx,ref} - P_{Tx,ref} \end{cases} \quad (S12)$$

Equation (S12) is precisely the origin of transmission. Therefore, the realized gain of the TLWA can be expressed as:

$$G_{TLWA} = G_{HA} + S_{21,TLWA} - S_{21,ref} \quad (S13)$$

For general commercial horn antennas, we may only know the far-field realized gain provided by the manufacturer. If we use this data directly, we will not be able to obtain the correct realized gains of the TLWA at all distances. Based on the calculation of the near-field gain expression of horn antennas in previous works [S2, S3], we can obtain the realized gains of the conical-horn antenna at all distances in an actual electromagnetic environment:

$$G_{HA}(d) = 10\lg\left\{\frac{4\pi^2 R_0^2}{\lambda^2}\left|\sum_{i=0}^{\infty}\frac{(-1)^i}{\left[(i+1)!\right]^2}\left(\frac{\pi R_0^2}{\lambda d}\right)^{2i}\right|\right\} - L \quad (S14)$$

In (S14), $L$ is used to correct the ideal assumption of the uniform aperture field. According to (S13) and (S14), the accurate realized gains of the TLWA at all distances and the conclusive expression used to combine theory and experiments to demonstrate the difference between near-field and far-field realized gains are presented in the main text.

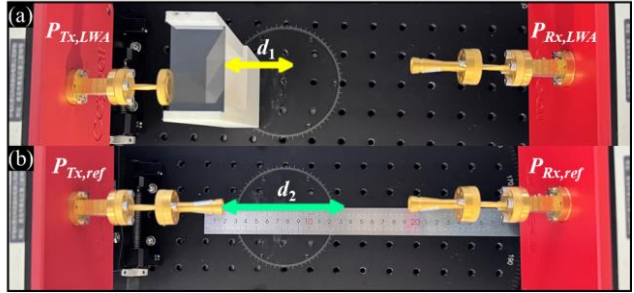

**FIGURE S6** | Schematic diagram of the realized gain of the proposed TLWA measured by comparative law. (a) Use the TLWA as the transmitting antenna and the horn antenna as the receiving antenna. (b) Both the transmitting antenna and the receiving antenna use horn antennas.

## S6 | Detailed Measured Results for Verification of the SLSR

To experimentally verify the side-lobe suppression ratio (SLSR) of the proposed TLWA in a practical terahertz environment, we measured its far-field normalized radiation pattern at $\theta = 80°$ over $\varphi$ ranging from $-90°$ to $90°$, as illustrated in Figure S7a. All measurements were performed under far-field conditions for both the TLWA and the horn antenna. Based on geometric relations, a $10°$ angular separation in the $\theta$ direction between the TLWA and the horn antenna was ensured. Figure S7b presents the measured far-field normalized radiation patterns ($\theta = 80°$) of the TLWA at different frequencies. It can be observed that the worst-case SLSR occurs at $\varphi = 60°$, yet even at this angle the suppression level exceeds 15 dB across all measured frequencies. These results confirm the clean single-beam characteristic of the TLWA and provide direct experimental evidence for the effectiveness of BA-type reshaping in suppressing side lobes.

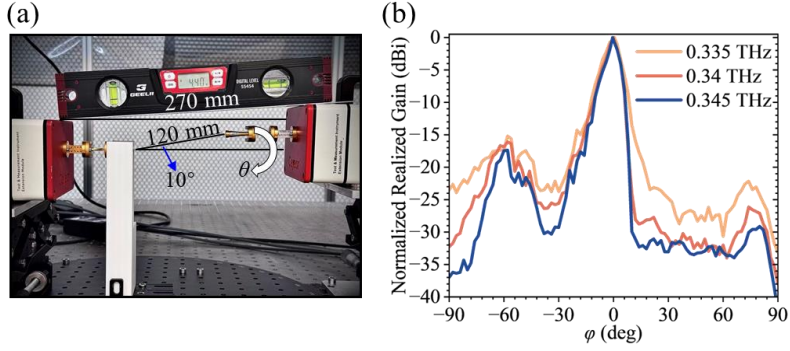

**FIGURE S7** | Presentation of the experimental setup and results of the proposed TLWA at $\theta = 80°$. (a) Presentation of the photograph of the TLWA and the equipment and methods required for measuring the far-field normalized radiation patterns at $\theta = 80°$. (b) Measured far-field normalized radiation patterns ($\theta = 80°$) of the TLWA at different frequencies.

## S7 | Detailed Simulated and Measured Far-Field Normalized Radiation Patterns

To further demonstrate the low-dispersion characteristics of the proposed TLWA, Figure S8 shows the simulated and measured normalized radiation patterns of $\varphi$ within a small range. It can be seen that as the frequency increases, both the simulated and measured main lobes shift toward the direction of negative $\varphi$ values. Due to subtle differences between the simulated and measured electromagnetic environments, there may be variations in the main-lobe deviation, but across the entire operating band, the maximum deviation of the main lobe can still be maintained within 2°.

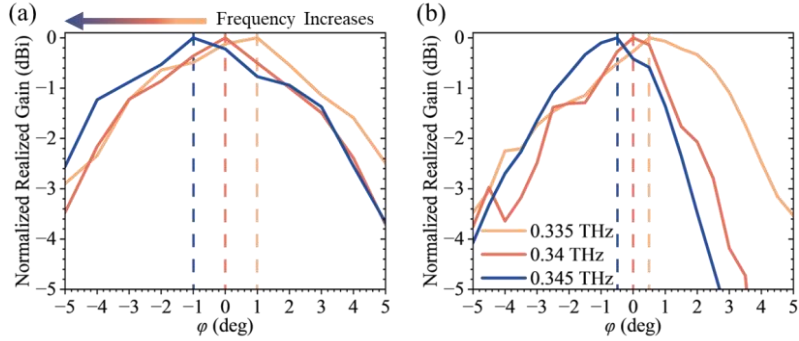

**FIGURE S8** | Detailed simulated and measured far-field normalized radiation patterns. (a) Simulated normalized radiation patterns ($\theta = 90°$, $-5° < \varphi < 5°$) of the TLWA. (b) Measured normalized radiation patterns ($\theta = 90°$, $-5° < \varphi < 5°$) of the TLWA.

## S8 | Efficiency Analysis of the Proposed TLWA

The total efficiencies and radiation efficiencies of the proposed TLWA are simulated, with results presented in Figure S9. The relationship between total efficiency and radiation efficiency is given by:

$$\eta_t = \left(1 - |\Gamma|^2\right)\eta_r \tag{S15}$$

In (S15), $\Gamma$ is the normalized reflection coefficient, $\eta_t$ and $\eta_r$ represent the total efficiency and the radiation efficiency, respectively. Figure S9a shows the efficiencies of the TLWA with larger structural footprint, while Figure S9b shows the efficiencies of the TLWA with smaller structural footprint. It is observed that both designs exhibit comparable total efficiency. In practice, a design trade-off exists between radiation efficiency and impedance matching: a slight reduction in radiation efficiency can lead to improved reflection performance, thereby bringing radiation efficiency closer to total efficiency. The TLWA with smaller structural footprint achieves near-unity radiation efficiency, yet may exhibit compromised reflection characteristics. This trade-off is absent in the TW, which operates solely on waveguiding without radiation. Consequently, the structural footprint can be optimized to balance compactness for on-chip integration with the performance requirements for both guided-wave and leaky-wave operation. As a result, the TLWA achieves an average total efficiency of 72%, while its radiation efficiency can be tailored between 72% and 100% flexibly by adjusting the structural footprint. In practical engineering, the reflection coefficient is often a more critical parameter for system performance. Therefore, in fabricating the TLWA, we intentionally adopted a larger footprint to prioritize impedance matching, accepting a reduction in radiation efficiency to achieve superior overall device performance.

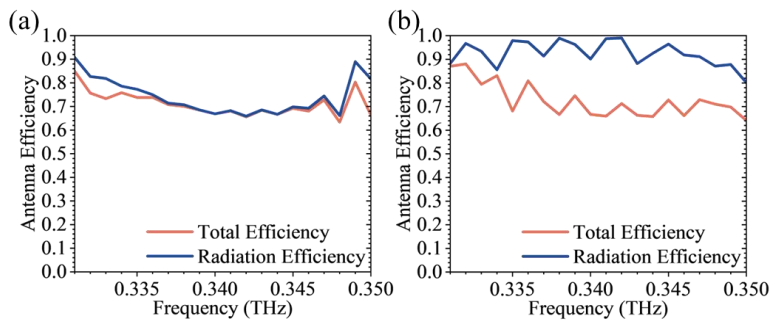

**FIGURE S9** | Efficiencies of the proposed TLWA. (a) Total efficiency and radiation efficiency of the TLWA with larger structural footprint. (b) Total efficiency and radiation efficiency of the TLWA with smaller structural footprint.

## S9 | Supplementary Terahertz Communication Experiment

Figure S10 provides further evidence of the unique value of the proposed TLWA as a receiving antenna for terahertz communication. As shown in Figure S10a, the transmitting horn antenna and the TLWA are replaced by two rectangular metal waveguides while keeping the rest of the system configuration identical to that in Figure 7 (2.5 mm free space, quadrature phase shift keying (QPSK) modulation format, 24 Gbps data rate, 0.338 THz center frequency). Under this condition, even at maximum optical power, the constellation diagram remains chaotic, indicating that no signal demodulation is possible (Figure S10b). This comparative experiment conclusively demonstrates that the TLWA possesses excellent radiation and reception capabilities, and its performance in the free-space communication link is constrained by the link budget rather than by the device itself. Figure S10c supplements the optical power spectrum before injecting into the uni-traveling-carrier photodiode (UTC-PD), the center frequency of the optical carrier $f_1$ is 193.548 THz and the frequency of the optical monophonic signal $f_2$ is 193.21 THz.

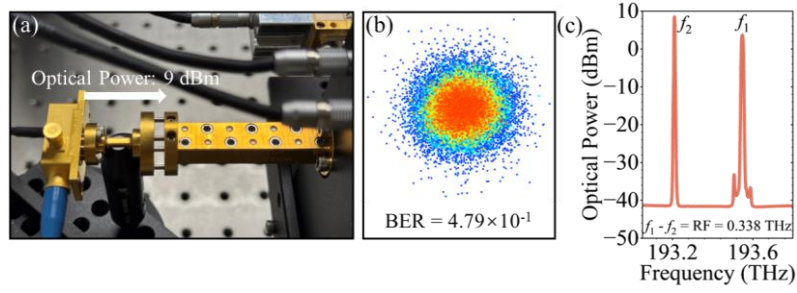

**FIGURE S10** | Supporting information of the terahertz communication experiment. (a) Photograph of the terahertz communication experiment with two straight waveguides spaced 2.5 mm apart. (b) Constellation diagram under 24 Gbps QPSK modulation format at optical power of 9 dBm. (c) Optical power spectrum before injecting into the UTC-PD.

## S10 | Comparison of the Proposed TLWA with Previous Works

Figure S11 shows the measured transmissions at different distances when the horn antenna is used for transmitting antenna and the proposed TLWA is used for receiving antenna. At identical distances, the transmission amplitudes measured in our setup are approximately 5 dB higher than those reported in reference [S1] (antenna realized gain: 12.2 dBi) and comparable to those in reference [S5] (antenna realized gain: 17 dBi). It should be noted that this improvement benefits from the use of a high-gain horn antenna at the transmitter side. Nevertheless, under the current configuration, the overall communication metrics achieved in our system remain below those reported in the aforementioned references. This indicates that a direct comparison of transmission distance does not fairly reflect the intrinsic performance of the antenna itself, as the achievable distance is primarily determined by the system's ability to compensate for link loss. Therefore, rather than emphasizing a direct comparison of transmission distance, this work focuses on demonstrating the functional capability and potential of the TLWA as a receiving antenna under given link conditions.

Table S1 compares the proposed TLWA with various terahertz passive antennas. It can be seen that end-fire antennas (orange-shaded rows) universally rely on tapers or graded refractive index buffers (GRIB) to achieve directional single-beam radiation, which inevitably increases fabrication complexity. In contrast, most leaky-wave antennas (blue-shaded rows) remain at the simulation stage due to fabrication and experimental challenges. Moreover, conventional leaky-wave antennas suffer from inherent dispersion, making it difficult to maintain stable directional radiation and resulting in limited effective bandwidth. Although reference [S8] employs the same terahertz platform as this work, its design objective is to achieve full-space radiation, which is fundamentally opposite to the directional single-beam goal pursued here. Consequently, its suitability for high-data-rate directional links may be limited. Additionally, the direct-coupling structure based on the AB-type valley-locked edge state in reference [S8] requires an extremely narrow defect width, making it prone to breakage during fabrication and less scalable toward higher frequencies such as 1 THz. In summary, although the proposed TLWA falls into the category of leaky-wave antennas, it combines the broad bandwidth and low dispersion typically associated with end-fire antennas, while offering large fabrication tolerance. These advantages make the proposed TLWA an ideal candidate for low-loss, highly integrated terahertz on-chip networks.

TABLE S1 | Performance Comparison of the Proposed TLWA with Various Terahertz Passive Antennas

| Ref. | Terahertz Platform | Fabrication Difficulty | Experiment | Mechanism | Directional | Frequency (THz) | EBW (%) | ARG (dBi) | DDR (Gbps) |
|---|---|---|---|---|---|---|---|---|---|
| [S4] (2018) | Silicon PC | Moderate | Yes | Taper Array | Yes | 0.343 | 21 | ~20 | 10 |
| [S5] (2025) | Silicon VPC | Moderate | Yes | GRIB + Berry Curvature Modification | Yes | 0.307 | ~3.3 | <17 | 20 |
| [S6] (2023) | Silicon SSPP | High | No | Leaky Mode | No | 0.735 | ~1.5 | ~12 | - |
| [S7] (2025) | Metal + Silicon VPC | High | No | Side Leakage + Horn Antenna | No | 0.403 | ~2.5 | ~20 | - |
| [S8] (2026) | Silicon VPC | Moderate | Yes | Leaky Edge State | No | ~0.32 | ~1.8 | <15 | 24 |
| **This Work** | **Silicon VPC** | **Low** | **Yes** | **Guided-Edge-State Leakage** | **Yes** | **0.3405** | **5.6** | **11** | **24** |

Note: PC is photonic crystal. VPC is valley photonic crystal. SSPP is spoof surface plasmon polariton. GRIB is graded refractive index buffer. EBW is effective bandwidth (absolute bandwidth/center frequency). ARG is average realized gain. DDR is directional data rate.

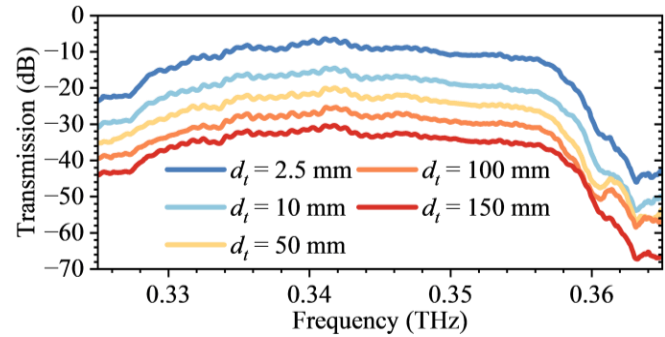

**FIGURE S11 |** Measured transmissions at different distances when the horn antenna is used for transmitting antenna and the proposed TLWA is used for receiving antenna.

## S11 | Fabrication Results of the Devices

Figure S12 shows the photographs of fabricated devices imaged by optical microscope. It can be seen that the structures of devices in each region are intact, with no defects caused by obvious processing errors. According to the main text, fabricated devices have three types of equilateral triangle air holes of different dimensions. In theory, their dimensions are 157.625 μm, 97 μm, and 84.875 μm, respectively. The actual fabrication results show that their dimensions are approximately 157.07 μm, 97.39 μm, and 84.82 μm, respectively. The above results indicate that the maximum fabrication error is only 0.4%. However, the wafer thickness error can be as high as 5%, which may cause frequency deviation of devices, but it will not have significant impacts on the other performances of devices.

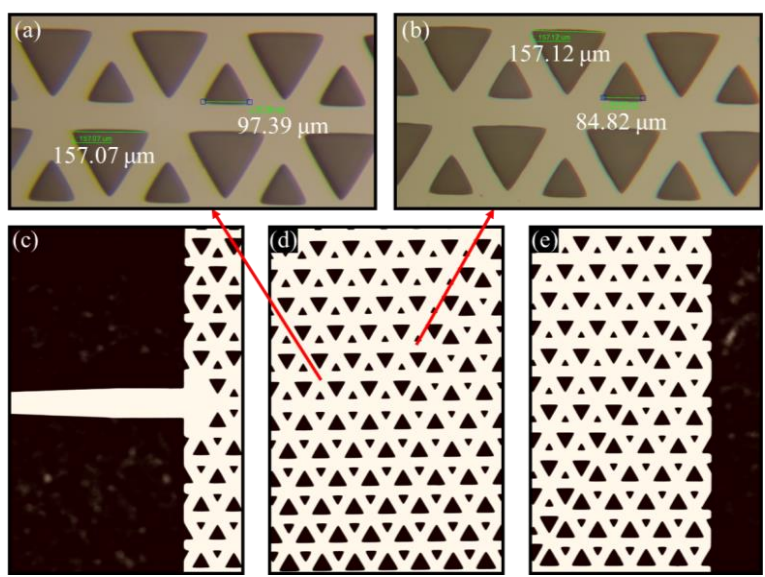

**FIGURE S12** | Photographs of the fabricated devices imaged by optical microscope. (a) Enlarged guided-wave section of the fabricated devices. (b) Enlarged leaky-wave section of the fabricated devices. (c) Taper coupler and guided-wave section coupling demonstration of the fabricated devices. (d) Guided-wave section and leaky-wave section coupling demonstration of the fabricated devices. (e) Leaky-wave section and air coupling (guided-edge-state leaky region) demonstration of the fabricated devices.

**References:**

S1. R. Jia, S. Kumar, T. C. Tan, et al., "Valley-conserved topological integrated antenna for 100-Gbps THz 6G wireless," *Science Advances* 9, no. 44 (2023): eadi8500.

S2. L. Xiao, Y. Xie, P. Wu, and J. Li, "Near-Field Gain Expression for Aperture Antenna and Its Application," *IEEE Antennas and Wireless Propagation Letters* 20, no. 7 (2021): 1225-1229.

S3. L. Xiao, Y. Xie, J. Li, and P. Wu, "Near-Field Gain Expression for Tapered Circular Aperture Antennas," *IEEE Transactions on Antennas and Propagation* 71, no. 9 (2023): 7684-7689.

S4. W. Withayachumnankul, R. Yamada, M. Fujita, and T. Nagatsuma, "All-dielectric rod antenna array for terahertz communications," *APL Photonics* 3, no. 5 (2018): 051707.

S5. S. Kumar, R. Jia, Y. J. Tan, et al., "Topological Berry Antenna on a Silicon Chip for Terahertz Wireless Communication," *Advanced Photonics Research* 6 (2025): 2500123.

S6. Y. Bai and S. Li, "Terahertz dual-beam leaky-wave antenna based on composite spoof surface plasmon waveguide. *Optoelectronics Letters* 19, no. 2 (2023): 72–76.

S7. H. Wang, H. Shi, W. E. I. Sha, et al., "Terahertz high-gain leaky-wave antenna utilizing valley topological photonic crystals with line defects," *Journal of Physics D: Applied Physics* 58 (2025): 215104.

S8. W. Wang, Y. J. Tan, P. Szriftgiser, G. Ducournau, and R. Singh, "On-chip topological leaky-wave antenna for full-space terahertz wireless connectivity," *Nature Photonics* 20 (2026): 317-323.